# A non-dispersive Raman D-band activated by well-ordered interlayer interactions in rotationally stacked bi-layer Graphene

Awnish K. Gupta[1*], Youjian Tang[1*], Vincent H. Crespi[1,2,3] and Peter C. Eklund[1,2,3#]

[1]*Department of Physics*
[2]*Department of Materials Science and Engineering*
[3]*Materials Research Institute*
[*] Both authors contributed equally to this work
*The Pennsylvania State University*
*University Park, Pennsylvania 16802 USA*

## Abstract

Raman measurements on monolayer graphene folded back upon itself as an ordered but skew-stacked bilayer (*i.e.* with interlayer rotation) presents new mechanism for Raman scattering in $sp^2$ carbons that arises in systems that lack coherent AB interlayer stacking. Although the parent monolayer does not exhibit a D-band, the interior of the skewed bilayer produces a strong two-peak Raman feature near 1350 cm$^{-1}$; one of these peaks is non-dispersive, unlike all previously observed D-band features in $sp^2$ carbons. Within a double-resonant model of Raman scattering, these unusual features are consistent with a skewed bilayer coupling, wherein one layer imposes a weak but well-ordered perturbation on the other. The discrete Fourier structure of the rotated interlayer interaction potential explains the unusual non-dispersive peak near 1350 cm$^{-1}$.

[#] Deceased.

Address correspondence to vhc2@psu.edu

One canonical feature of the Raman response of all sp$^2$ carbons is the so-called D-band near 1350 cm$^{-1}$. Whereas most Raman scattering requires a near-zero phonon wave-vector to conserve crystal momentum, the D-band involves a finite-wavevector phonon. Therefore it is activated by disorder and hence provides a powerful, universal means to diagnose the degree of structural order in all sp$^2$ carbons: nanotubes, graphite, graphene and nanoporous carbons. Here we demonstrate a new mechanism for the generation of a Raman D-band, one which does *not* require structural disorder. Instead, the translational symmetry breaking arises from an interlayer rotation between two well-ordered graphene sheets. The usual disorder-dependent D-band peak shifts in energy as the laser wavelength is varied, because many different phonon wavevectors can resonate against the broad spatial Fourier spectrum of localized structural disorder. In contrast, the new interlayer Raman "I-band" is pinned to specific, discrete Fourier components of the interlayer potential, hence it contains prominent *non-dispersive* components. It is essentially Bragg scattering, as measured by Raman.

Graphene multilayers that are prepared [1] by transfer from slabs of graphite are presumably AB stacked as in the parent material, with half the atoms in one layer sitting above hexagon centers in the layer below [2]. Well-ordered interlayer stacking that deviates from the standard AB and ABC geometries is much more difficult to produce, and the properties of these systems are not well-known. Here we demonstrate that the accidental folding of a graphene monolayer (ML) back onto itself at a skewed (i.e. non-AB) angle produces a unique Raman signature that reveals a weak, effectively incommensurate interlayer interaction. Few-layer graphene flakes were prepared by micro-mechanical cleavage of highly oriented pyrolitic graphite [1]. These *n*-layer graphene (*n*GL) flakes sometimes fold back upon themselves [3-5]; a partial fold provides a convenient plateau across which to count the number of layers in AFM [6]

– here we focus on the case of a folded monolayer. Since the fold forms in a singular, unannealed mechanical event, the lattice registry across it need not respect AB stacking. Hence, it provides a means to study alternative stackings between layers, either higher-order commensurate interlayer rotations or fully incommensurate stacking. These folded/stacked regions can be directly compared to monolayer regions of the *same* parent flake. Raman spectra from such folds were collected with single-grating (Renishaw InVIA) and triple-grating (JY T64000) Raman microscopes equipped with cooled CCD cameras, calibrated by standard atomic emission lines, and excited by Ar-Kr ion (Coherent Innova 70C). First- and second-order Raman spectra can easily distinguish a graphene monolayer from a commensurate bilayer (CBL) [7-9] that may form when, for example, a preexisting AB-stacked bilayer is cleaved from the graphite source. The CBL has a four-peak structure for the two-phonon 2D (also called G′) band due to band splitting from the interlayer interaction [9], whereas the ML has a single 2D peak. Also, the G-band of the CBL is broader than that of the ML, and the G-band intensity in commensurately stacked $n$GLs increases linearly with the number of layers [7, 8] for fixed excitation frequency. An interlayer-rotated bilayer (IBL) also has a distinct Raman signature. Both the G-band and the 2D band of the IBL are at least as narrow as those of the ML, as reported elsewhere [10]: this suggests an unusually weak interlayer interaction in the IBL, as compared to the large band splitting of the CBL. However, the most striking distinguishing feature of the IBL Raman spectrum is a band near ~1352 cm$^{-1}$ (at 514.5 nm excitation), which is absent from the interiors of the ML and CBL regions of the same sample(s), as discussed below.

Figure 1 shows the Raman response in the spectral region from 1300 to 1420 cm$^{-1}$ for five different skewed (*i.e.* with interlayer rotation) bilayer samples (S1 to S5) at 514.5 nm laser excitation. Spectra are offset vertically for clarity and fitted with two Lorentzian peaks. The

~1350 cm$^{-1}$ Raman band in the SkBL contains distinct peaks: a stronger, slightly asymmetric component that we call $I_{envelope}$, which appears around $\omega_{envelope}$~1352 cm$^{-1}$, and a second peak (called $I_{fixed}$ for reasons to become clear shortly) which appears at different frequencies in different IBLs ($\omega_{fixed}$~1384, 1382, 1371, 1384 and 1395 cm$^{-1}$ in S1, S2, S3, S4 and S5, respectively). Scanned Raman measurements show that the overall center frequency of the I-band is nearly constant across the interior of the IBL, so the interaction that gives rise to this Raman response is not an edge effect and is homogeneous across long length scales [10]. In addition, the monolayer regions of the same flake do not show any Raman features in the range 1300 – 1400 cm$^{-1}$, hence the parent flake does *not* contain significant in-plane disorder and this new feature must arise somehow from the interlayer stacking. We call $I_{envelope}$ & $I_{fixed}$ collectively the "I-band," where "I" refers to "interlayer" and is also suggestive of "incommensurate". In all likelihood, these five samples have several different rotation angles between the constituent layers. The larger scatter for $\omega_{fixed}$ then suggests that the precise interlayer rotation angle affects $\omega_{fixed}$ more strongly than $\omega_{envelope}$. The upper right corner of each spectrum in Figure 1 shows the intensity of $I_{envelope}$ relative to the G-band: this ratio varies greatly, from 0.02 to 0.1. In the spectra with highest signal-to-noise, (i.e. S1, S2 and S4), $I_{fixed}$ is significantly sharper than $I_{envelope}$. All these spectral features are explained by the model given below.

Figure 2 shows Raman spectra from 1300 to 1400 cm$^{-1}$ for five different laser excitation wavelengths: 363 nm, 457 nm, 488 nm, 514 nm and 647 nm, as collected from sample S1. The spectra are fit with two Lorentzian peaks ($I_{envelope}$ and $I_{fixed}$), scaled by the $I_{envelope}$ intensity, and plotted with a vertical offset. The peaks $I_{envelope}$ and $I_{fixed}$ clearly behave very differently as a function of laser excitation energy. $I_{envelope}$ disperses at $d\omega_{envelope}/dE_{laser}$ ~ 49 cm$^{-1}$/eV, while $I_{fixed}$ show essentially no dispersion: $d\omega_{fixed}/dE_{laser}$ ~ -2 cm$^{-1}$/eV, hence the name "fixed". In addition, the relative intensity $I_{fixed}/I_{envelope}$ varies rapidly with changing laser energy. The overall I-band intensity ($I_{envelope}+I_{fixed}$) increases with increasing laser excitation [10], a trend opposite to that observed for D-bands in other sp$^2$ carbon systems, *i.e.* graphite [11], glassy carbon [12], nanocrystalline graphite [13], and carbon nanotubes [12]. Finally, the shortest-wavelength laser excitation, 363 nm, produces a second sharp peak, labeled $I_{fixed}(2)$.

These unusual Raman features occur in the same spectral range as the D-band of disordered sp$^2$ carbons. To best understand its origin, we begin with a brief recapitulation of the standard double-resonant D-band process. The D-band arises when a finite-wavevector mode becomes Raman active in the presence of in-plane disorder such as vacancies, pentagon-heptagon pair defects, or finite basal plane dimensions that break translational invariance [14-17]. Since the apexes of the conical π (filled) and π· (empty) bands of monolayer graphene touch at the corners K and K′ of the hexagonal Brillouin zone, the optical transitions involved in resonant Raman scattering define a roughly circular loci of electronic states (with some trigonal distortion) around K and K′, whose radii depend on the photon energy E. Under the double-resonant process active in the D-band [15], this electron absorbs a photon (1→2) and scatters across the Brillouin zone to the vicinity of the other K′/K point (2→3, as shown by the arrow labeled *q* to the left of the Γ point in Figure 3), then scatters back to near the original location (3→4) before de-exciting through emission of a Raman-shifted photon (4→1). One scatter is mediated by a phonon $\omega_q$, the other by a static disorder potential $V_{-q}$:

$$M_{DR} = \sum_{2,3,4} \left[ \frac{M_{1\to2} M_{2\to3} M_{3\to4} M_{4\to1}}{(E_1 - E_2)(E_1 - \hbar\omega_{ph} - E_3)(E_1 - \hbar\omega_{ph} - E_4)} \right]. \quad (1)$$

Contributions to this process from across the Brillouin zone are modulated by the resonant denominators of equation (1), hence the term double resonance. The overall D-band response is dominated by the maximal and minimal phonon wave-vectors that span the electronic loci, since these define sharp maxima in the reciprocal-space density of resonant transitions [18] while the Fourier signature of localized in-plane disorder (such as a vacancy or layer edge) is broad and featureless in reciprocal space. The resulting D-band in normal disordered $sp^2$ carbons is somewhat broad [19], about 15 cm$^{-1}$ full-width-at-half-maximum. Since the lengths of the maximal and minimal wavevectors for the double-resonant transition vary with laser excitation frequency, the standard D-band always disperses with laser energy at about ~50 cm$^{-1}$/eV [20].

The IBL region of each sample S1-S5 is made from the same sheet as the ML region, so it shares similar in-plane disorder. Why then does the IBL show a Raman response in the D-band region, when the ML portion of the same flake does not? First, note that Raman features in the D-band range *do not* require disorder-- broken translational invariance is enough. Crucially, whereas $I_{envelope}$ disperses by the expected ~50 cm$^{-1}$/eV under variable photon excitation energy, $I_{fixed}$ shows *essentially no dispersion* within experimental errors: ~ -2 cm$^{-1}$/eV in S1. To our knowledge, this is the first non-dispersive Raman feature seen near 1350 cm$^{-1}$ in $sp^2$ carbon. Since the interiors of the ML and CBL regions of the same flake have no features near 1350 cm$^{-1}$, the $I_{envelope}$ and $I_{fixed}$ peaks in the IBL cannot arise from in-plane disorder. Instead, they apparently arise from the rotated interlayer stacking of the two layers, where one layer of the IBL acts as a weak perturbation on the other. Consider double-resonant Raman scattering within the lower layer of an IBL, where the phonon momentum is compensated not by localized in-plane disorder, but by the *well-ordered, periodic* (but rotated) static perturbing potential of the upper layer. The basic mechanism is the same as discussed above for the D-band, but arbitrary crystal momentum is no longer available from the static potential. Fig. 3 depicts the resulting situation in reciprocal space. The hexagons show the extended zone scheme for the lower layer (the extended zone scheme most easily shows Umklapp processes). The light blue disc depicts, schematically, one of the regions of allowed double-resonant phonon wavevectors for a given photon energy: these are all wavevectors that span the two circular loci depicted schematically to the left of the Γ point. (Similar discs appear around each repeated-zone corner, but only one is shown for clarity). The dark blue dots provide the critical new geometrical element of the IBL: they are the Fourier components of the static potential of the upper layer, considered as a perturbation on the lower layer. They are rotated by θ with respect to the grey dots, which are the equivalent points for the lower layer. For an Umklapp defined by *G*, the Fourier component *$G_\theta$* of the interlayer perturbation supports double-resonant scattering at phonon wavevector *q*. As a function of the rotation angle between the layers, different Fourier components of the interlayer perturbation can support double-resonant Raman transitions. Crucially, a given interlayer-activated Raman transition, defined by *$G_\theta$*, does *not* disperse with laser photon energy, since it is pinned to a specific wave-vector of the interlayer perturbation.

To demonstrate that the interlayer interaction is strong enough to induce significant Raman scattering, we have performed density functional theory (DFT) calculations for a rotated bilayer, using the projector-augmented waves in the local density approximation [21] (LDA) with a plane wave basis energy cutoff of 400 eV in the Vienna Ab-initio Simulation Package [22-26]. The upper and lower layers are skewed by $\cos^{-1}(37/38)$, which yields a computationally tractable 76-atom periodic supercell. The essential physics of this order-of-magnitude estimate are independent of the precise value of the rotation angle between layers. The DFT/LDA equilibrium interlayer spacing, 0.34 nm, is slightly larger than that calculated for commensurate

AB stacking, 0.32 nm, due to a less efficient interlayer registry. Both values are smaller than experiment, due to the standard LDA overbinding. The weak perturbative nature of the interlayer interaction has already been established [10, 27, 28]. To establish that the static interlayer perturbation is sufficiently strong to support non-neglible Raman transitions, we calculate the self-consistent crystal potential for the rotated bilayer and then subtract off the self-consistent crystal potential of the lower ML alone to obtain the perturbation on the lower ML from the upper ML. We have also obtained the perturbation on the lower ML by simply calculating the crystal potential of an isolated upper ML at the position of the lower ML; the two methods give very similar results. This weak perturbation (from either method) is Fourier decomposed on the (rotated) reciprocal lattice of the upper layer. The Fourier components of the perturbation potential decay quickly with increasing wavevector: only the smallest three or four Fourier components are significant. For example, in a plane parallel to the graphene layers but 0.9 Angstroms below from the upper (perturbing) layer, the ampltudes of first 4 Fourier components decrease in the ratio (1, 0.52, 0.17, 0.02), where the first Fourier component has been normalized to 1. We estimate the order-of-magnitude of a typical $I_{fixed}$ transition by integrating the calculated interlayer perturbation against two single-layer electronic states whose wave vectors satisfy the resonance condition. The result, ~0.15 meV, is about 1000 times smaller than a typical electron-phonon matrix element between similar states [29, 30], which is consistent with the experimentally measured relative peak intensities of the I-band and the main Raman peaks: experimentally, the I-band is much weaker than the main Raman peaks.

How does rotated interlayer geometry affect the overall double-resonant response? The shape and dispersion of the standard D-band are governed by the distribution of resonant electronic transitions in reciprocal space. The I-band adds a new element: the sharpness of the static scattering potential in reciprocal space. Consider first an over-simplified Gendanken scenario where all Fourier components of the static interlayer perturbation have equal amplitude. Contributions from the rotated blue lattice of *G$_θ$* in Figure 3 would extend to infinite distance from the origin, and for any choice of interlayer rotation angle many different Fourier components of the static potential (blue dots) could compensate for double-resonant phonons (i.e. fall within light blue disks). Since many different wavevectors would contribute, weighed according to the standard double-resonance conditions, something similar in shape to a standard *dispersive* D-band would result. Now relax the Gedanken requirement that all Fourier components of the static interlayer perturbation be equal in strength: the I-band becomes *discrete*, with contributions arising from phonons that match to distinct, substantial Fourier components of the static interlayer perturbation. A double-resonant transition mediated by a particularly strong Fourier component $V_q$ will appear at a *fixed* non-dispersive frequency. Such non-dispersive peaks will appear at different frequencies for different interlayer rotation angles between layers, hence the appearance of $I_{fixed}$ at 1384, 1382, 1371, 1384 and 1395 cm$^{-1}$ in different IBLs. When the laser excitation frequency is changed, the frequencies of the component peaks within the I-band *do not change*, since they are pinned to the Fourier structure of the interlayer perturbation. Instead, their amplitudes evolve to reflect the resonant weights of these double-resonant wave-vectors at the new photon energy. If many different, individually unresolved Fourier components contribute, then they may sum to produce a broad asymmetric band whose overall envelope appears to disperse with laser energy is a manner similar to the standard D-band. Hence we attribute an $I_{fixed}$ peak to a particularly prominent Fourier component of the interlayer perturbation, and $I_{envelop}$ to the residual unresolved contributions from the weaker Fourier components. (As a caveat, note that we cannot rule out that $I_{envelop}$ actually arises from some hidden source of disorder that is only present in the folded bilayer, such as *e.g.* trapped adsorbates underneath the fold. Of course, the new Raman feature of present focus, $I_{fixed}$, cannot arise from such a source). This scenario predicts that more than one $I_{fixed}$ is possible, particularly at higher laser excitation energies where the allowed double-resonant regions in reciprocal space expand. In fact, at the shortest laser excitation wavelength, 363 nm, a new peak at 1410 cm$^{-1}$ is observed. Although the observed $I_{fixed}$ frequencies, plus knowledge of the phonon dispersion curves, provide some information on the magnitude of the corresponding phonon wavevector, they do not tightly constrain the direction of this phonon, so the available experimental information is not sufficient to determine the precise rotation angle between the layers. However, we can predict the qualitative evolution of the strength of $I_{fixed}$ with laser excitation. Experimentally, the strength of $I_{fixed}$ increases with increasing laser excitation energy. Making the standard assumption of constant matrix elements in the numerator of Eqn. (1), we have calculated the transition amplitude $M_{DR}$ for $\omega_{phonon}$ = 0.1675 eV on a 100 x 100 reciprocal space grid (interpolated to 3000 x 3000 by triangular interpolation), using the 76-atom supercell described earlier. Upon increasing the laser excitation energy from 2.5 eV to 3 eV, the variation in $M_{DR}$ is consistent with the experimental trend: it increases by factors of 2.4, 1.8 and 1.7 for $G_\theta$ = (0,2), (-2,1) and (0,3) respectively [i.e. the three highlighted $G_\theta$ of Fig. 3, each of which satisfies double resonance for $\theta = \cos^{-1}(37/38)$]. Finally, we note that nondispersive I-band features from fully incommensurate bilayers and commensurate systems with large, skewed supercells are essentially indistinguishable, since both can support these Fourier-pinned double-resonant transitions.

In summary, the optical response of a skew-stacked graphene bilayer reveals a new mechanism for Raman scattering in sp$^2$ carbons, an *interlayer band,* or I-band, wherein a well-ordered, periodic, perturbing potential arising from rotated interlayer stacking generates non-dispersive Raman features that are pinned to specific Fourier components of the interlayer perturbation. These features appear in a spectral region that was previously associated only with disorder. Since the I-band is constructed from just those phonons that match onto wave-vectors of the static interlayer perturbation, it provides an indirect measure of the Fourier transform of the graphite crystal potential, measured at a slight remove from the basal plane: is this sense it is Bragg scattering via Raman. Interestingly, an unusually sharp Raman peak at ~1333 cm$^{-1}$ (FWHM ~ 5 cm$^{-1}$) has been seen in strained suspended nanotubes [31], although its origin is unclear, and it is not known whether this strain-induced Raman peak disperses with laser energy.

**Acknowledgements:** This work was supported in part by the NSF NIRT program under ECS-0609243. We acknowledge assistance from Ke Zou in acquisition of AFM images.

**References**:

[1]   K. S. Novoselov, *et al.*, PNAS **102**, 10451 (2005).
[2]   J. W. Mcclure, Phys. Rev. **1**, 612 (1957).
[3]   Z. Ni, *et al.*, Phys. Rev. B **77**, 235403 (2008).
[4]   J. C. Meyer, *et al.*, Nature **446**, 60 (2007).
[5]   P. Poncharal, *et al.*, Phys. Rev. B **78**, 113407 (2008).
[6]   K. S. Novoselov, *et al.*, Science **306**, 666 (2004).
[7]   A. Gupta, *et al.*, Nano Lett. **6**, 2667 (2006)
[8]   D. Graf, *et al.*, Nano Letters **7**, 238 (2007).
[9]   A. C. Ferrari, *et al.*, Phys. Rev. Lett. **97**, 187401 (2006).
[10]  A. K. Gupta, *et al.*, (unpublished results).
[11]  Y. Wang, D. C. Alsmeyer and R. L. Mccreery, Chem. Mater. **2**, 557 (1990).
[12]  S. D. M. Brown, *et al.*, Phys. Rev. B **64**, 073403 (2001).
[13]  K. Sato, *et al.*, Chem. Phys. Lett. **427**, 117 (2006).
[14]  R. Saito, *et al.*, Phys. Rev. Lett. **88**, 027401 (2002).
[15]  C. Thomsen and S. Reich, Phys. Rev. Lett. **85**, 5214 (2000).
[16]  A. C. Ferrari and J. Robertson, Phys. Rev. B **61**, 14095 (2000).
[17]  A. K. Sood, R. Gupta and S. A. Asher, J. Appl. Phys. **90**, 4494 (2001).
[18]  J. Maultzsch, *et al.*, Phys. Rev. Lett. **92**, 075501 (2004).
[19]  A. K. Gupta, *et al.*, ACS Nano **3**, 45 (2009).
[20]  M. A. Pimenta, *et al.*, PCCP **9**, 1276 (2007).
[21]  J. P. Perdew and A. Zunger, Phys. Rev. B **23**, 5048 (1981).
[22]  H. J. Monkhorst and J. D. Pack, Phys. Rev. B **13**, 5188 (1976).
[23]  G. Kresse and J. Furthmüller, Phys. Rev. B **54**, 11169 (1996).
[24]  G. Kresse and J. Furthmüller, Comput. Mater. Sci. **6**, 15 (1996).
[25]  P. E. Blo¨chl, Phys. Rev. B **50**, 17953 (1994).
[26]  G. Kresse and J. Hafner, Phys. Rev. B **47**, 558 (1993).
[27]  J. M. B. Lopes dos Santos, N. M. R. Peres and A. H. Castro Neto, Phys. Rev. Lett. **99**, 256802 (2007).
[28]  S. Latil, V. Meunier and L. Henrard, Phys. Rev. B **76**, 201402 (2007).
[29]  G. D. Mahan, Phys. Rev. B **75 68**, 125409 (2003).
[30]  A. H. Castro Neto and F. Guinea, Phys. Rev. B **75**, 045404 (2007).
[31]  S. W. Lee, G. H. Jeong and E. E. B. Campbell, Nano Lett. **7**, 2590 (2007).

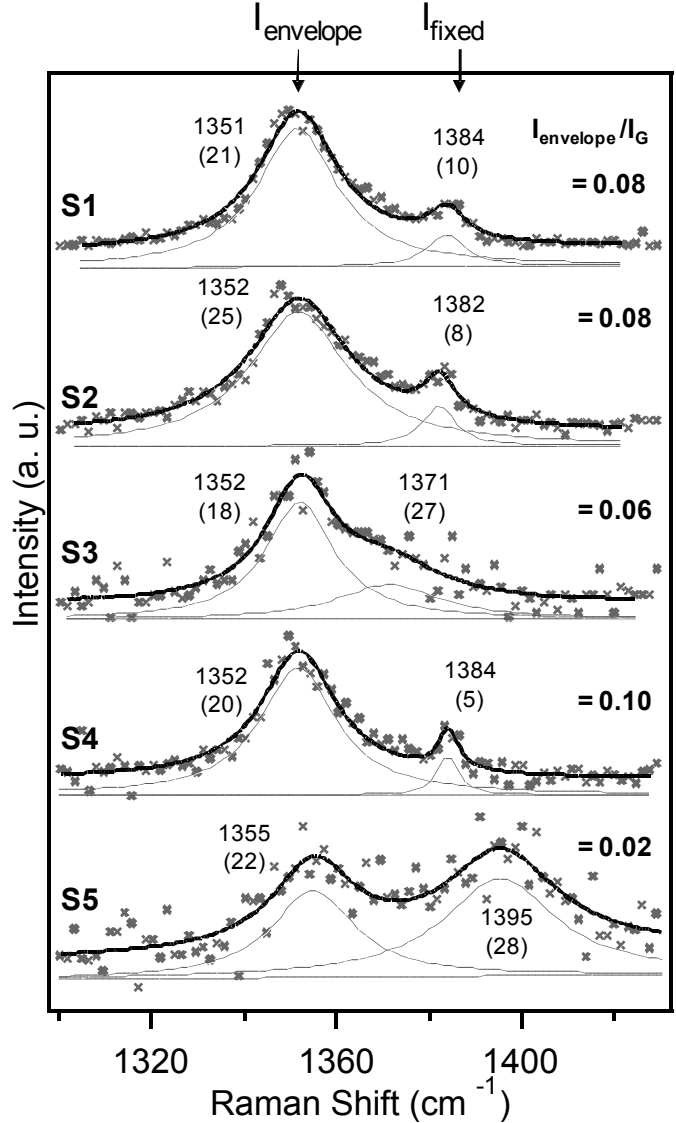

Figure 1: Raman spectra in the frequency range from 1300 to 1430 cm$^{-1}$ for five different skewed bilayer graphene films (i.e. with interlayer rotation), labeled S1 to S5. All spectra are collected using a 514.53 nm laser excitation at room temperature under ambient conditions. Peak positions and FWHM (in parentheses) are marked next to each Raman peak. The lower-frequency peak is called $I_{envelope}$ and the higher-frequency peak $I_{fixed}$, as discussed in the main text. $I_{envelope}/I_G$ is shown to the right of each spectrum.

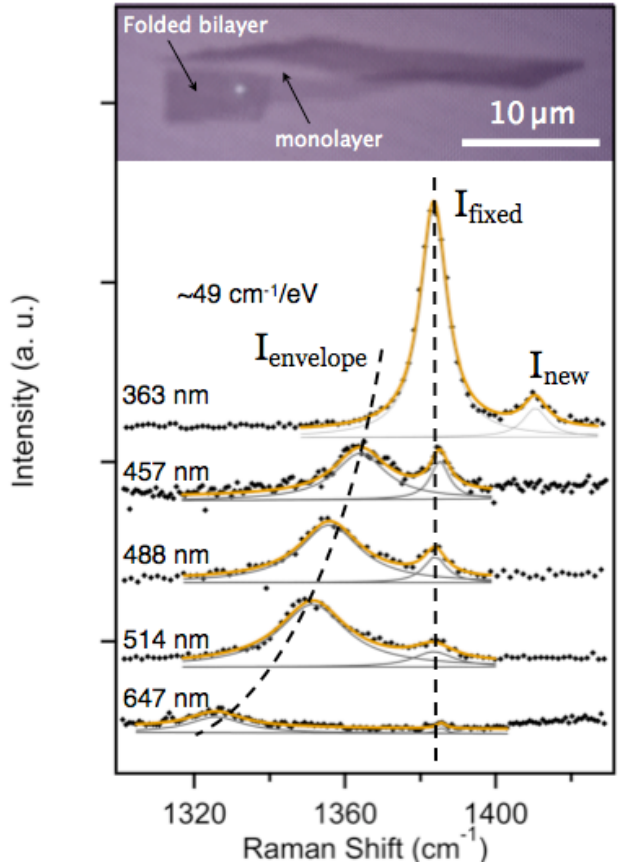

Figure 2: (a) Raman spectra in the range from 1300 to 1400 cm$^{-1}$ for excitation wavelengths of 363, 457, 488, 514 and 647 nm, showing the dispersion of $I_{envelope}$ and the unusual absence of dispersion for $I_{fixed}$. At the highest laser excitation energy, a new I-band peak emerges near 1410 cm$^{-1}$, as discussed in the text.

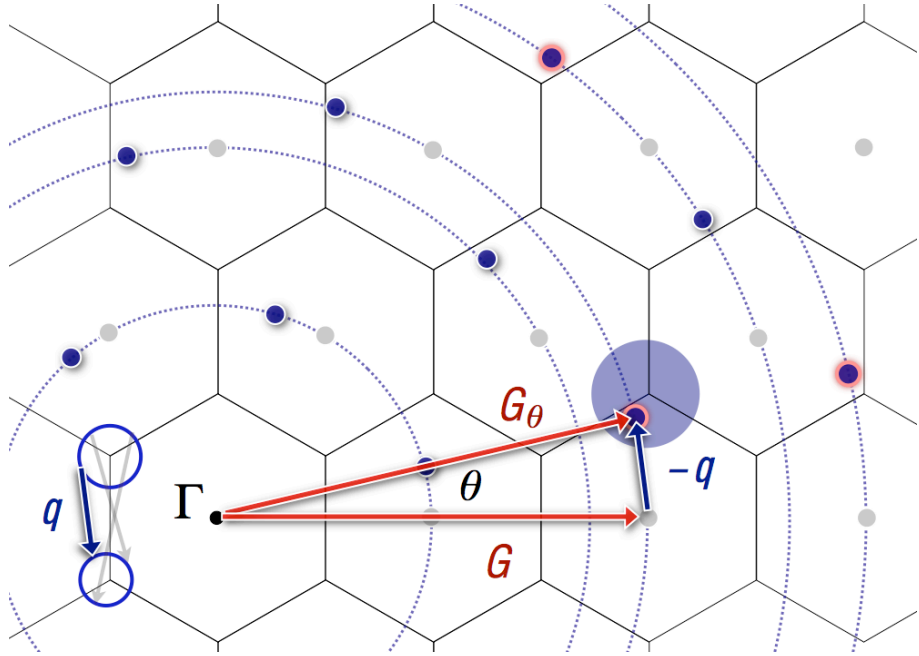

Figure 3: Schematic representation of the I-band scattering, whereby the static interlayer perturbation activates finite-wavevector double-resonant Raman scattering. The light blue disc represents one region that supports double-resonant Raman scattering, as defined by the double-resonant process depicted to the left of $\Gamma$ (similar discs surround every corner of the repeated zones). The dots represent Fourier components of the perturbing potential from the adjacent layer: grey dots describe commensurate AB stacking while dark blue dots describe a rotated stacking. In general, double-resonant I-band scattering (i.e. by phonons of wavevector $q$ whose wave-vector is compensated by a static interlayer perturbation at $V_{-q}$) occurs whenever a blue dot intersects a light blue disc.